# Status of $\theta_{13}$ measurement in reactor experiments

Kwong Lau
*University of Houston, Houston, TX 77204, USA*

The status of $\theta_{13}$ measurements from Daya Bay, RENO and Double Chooz experiments is reviewed.

## 1. INTRODUCTION

Experiments conducted in the last several decades with neutrinos from various sources have confirmed that the three types (flavors) of neutrinos have small but finite rest mass [1]. In the framework of three flavors, neutrinos oscillate in vacuum according to three mixing angles, two mass-squared differences, and one CP-violating phase angle. The unitary transformation relating the mass and flavor eigenstates is customarily parameterized by the Pontecorvo-Maki-Nakagawa-Sakata (PMNS) matrix, defining the three mixing angles ($\theta_{23}$, $\theta_{12}$, and $\theta_{13}$) and the CP-violating phase ($\delta$) [2],

$$U = \begin{pmatrix} 1 & 0 & 0 \\ 0 & c_{23} & s_{23} \\ 0 & -s_{23} & c_{23} \end{pmatrix} \begin{pmatrix} c_{13} & 0 & s_{13}e^{-i\delta} \\ 0 & 1 & 0 \\ -s_{13}e^{i\delta} & 0 & c_{13} \end{pmatrix} \begin{pmatrix} c_{12} & s_{12} & 0 \\ -s_{12} & c_{12} & 0 \\ 0 & 0 & 1 \end{pmatrix}$$

where $c_{23} = \cos\theta_{23}$ etc. Two of the mixing angles ($\theta_{12}$ and $\theta_{23}$) are large, and have been measured: $\sin^2\theta_{12}=0.306+0.018-0.015$ and $\sin^2\theta_{23}=0.42+0.008-0.005$ [3]. The mass-squared differences have also been measured in the so-called atmospheric and solar regimes, $\Delta m^2_{31} = (2.35 +0.12-0.09) \times 10^{-3}$ eV$^2$ and $\Delta m^2_{21} = (7.58+0.22-0.26) \times 10^{-5}$ eV$^2$, respectively [3].

The absolute rest masses of the neutrinos are still not known; only the magnitudes of the differences are known, an ambiguity referred to as the mass hierarchy problem in neutrino physics. As one can see from the PMNS matrix, amplitudes related to leptonic CP violation are proportional to $\sin\theta_{13}$. It is, therefore, of great importance to measure $\theta_{13}$ as precisely as possible. This may shed some light on the matter-antimatter asymmetry mystery of the universe, nominally considered associated with CP violating interactions. Earlier results from MINOS [4], T2K [5], and Double CHOOZ [6] suggest that $\theta_{13}$ can be large. The Daya Bay Experiment, based on 51 days of data with six antineutrino detectors (ADs), first published a measurement of non-zero $\theta_{13}$, with a significance of more than 5 standard deviations [7]. The large value of $\theta_{13}$ was subsequently confirmed by RENO [8] and Daya Bay with 2.5 times the data published in Reference 7 [9]. This talk reviews the methodology of reactor approach in $\theta_{13}$ measurement, and summarizes the current status of $\theta_{13}$ measurements. $\theta_{13}$ can also be measured in long-baseline accelerator neutrino experiments; an update on the most recent T2K results can be found in this proceedings [10].

## 2. REACTOR NEUTRINO EXPERIMENTS

### 2.1. Neutrino survival probability and detection

The present generation of reactor neutrino experiments was designed to explore the potentially small value of $\theta_{13}$ by measuring the survival probability of electron antineutrinos from the most powerful nuclear reactors. In nuclear reactors, the antineutrinos are derived from beta decays of nuclear fission products. The energies of reactor neutrinos are below the threshold for muon production. Hence, the only way to measure θ$_{13}$ is through the disappearance of the antineutrinos. The probability of an electron antineutrino of energy E (in MeV) remaining an electron antineutrino (survival probability) after traveling a distance of L (in meters) is given by

$$P_{sur} = 1 - \sin^2 2\theta_{13} \sin^2\left(1.267\frac{\Delta m^2_{31} L}{4E}\right) - \cos^4\theta_{13} \sin^2 2\theta_{12} \sin^2\left(1.267\frac{\Delta m^2_{21} L}{4E}\right). \quad (1)$$

The three main reactor experiments covered in this talk and their basic parameters are listed in Table 1. They all search for the disappearance of reactor neutrinos (neutrinos from here on refer to electron-type antineutrinos unless specified otherwise) in detectors located at a distance L ~ 2 km. At this baseline, the contribution from the first term in Equation 1, proportional to sin$^2$2θ$_{13}$, is maximal, and the contribution from the second term is negligible. It is noteworthy that the survival probability does not depend on the CP violating parameter δ. Furthermore, matter effects are negligible at this distance. The reactor neutrinos thus provide an unambiguous measurement of θ$_{13}$.

The reactor neutrinos are detected by the neutron Inverse Beta Decay (IBD) reaction in liquid scintillator,

$$\bar{\nu}_e + p \rightarrow n + e^+,$$

where free protons in the form of hydrogen constitute typically 10% of the mass of the liquid scintillator. The IBD reaction is characterized by a sequence of interactions, first the prompt signal coming from the production and subsequent annihilation of the positron, followed by the capture of the moderated neutron in the liquid scintillator. To suppress background, the liquid scintillator in the target volume is doped with about 0.1% by weight of Gadolinium (Gd). The large capture cross section of thermal neutrons by Gd reduces the capture time from the typical hydrogen capture time to about 30 μs. The capture of a neutron by Gd results in several gammas of about 8 MeV total energy. The IBD event topology is illustrated in Figure 1 (left panel). The IBD event rate is given by a convolution of the neutrino flux and the IBD cross section, both dependent on the energy of the neutrino:

$$\frac{N}{T} = \frac{\varepsilon N_p}{4\pi L^2} \int \Phi_\nu(E) \sigma_\nu(E) dE$$

The yield (N) in time T depends on the number of target protons (N$_p$), baseline (L), IBD detection efficiency (ε), neutrino flux (Φ$_\nu$), and the IBD cross section (σ$_\nu$). The dominant uncertainly is the neutrino flux which depends on the details of nuclear fissions in the reactor. Major uncertainties are the neutrino energy spectrum of each fission isotope, the fuel burn-up cycle, and many other factors. Both commercial and special computer programs are used to pin down the overall flux uncertainty, typically to 2-3%. Fortunately, most of these uncertainties are canceled when a ratio of IBD rates is considered, a strategy used by all θ$_{13}$ reactor experiments (see 2.1.1).

The IBD events are selected by applying cuts to the data, exploiting the timing and energy features of the signal events. The event selection criteria, using the Daya Bay experiment as an example, are:

- Prompt energy between 0.7 and 12 MeV;
- Delayed energy between 6 and 12 MeV;
- Muon veto: 0.6 ms to 1 s, depending on energy deposited in the neutrino detector, after the passage of a muon through the neutrino detector;

- Neutron capture time between 1 and 200 μs; and
- Multiplicity cut of no other event 200 μs before and after each IBD.

All the experiments suffer from some unexplained PMT flashes which can be removed by applying simple cuts without losing real events. These cuts result in a clean sample of IBD events with little background. The measurement of $\theta_{13}$ remains challenging due to large uncertainties in the neutrino flux and detector systematics. In order to achieve high sensitivity to $\theta_{13}$, the experiments were designed with the following key features.

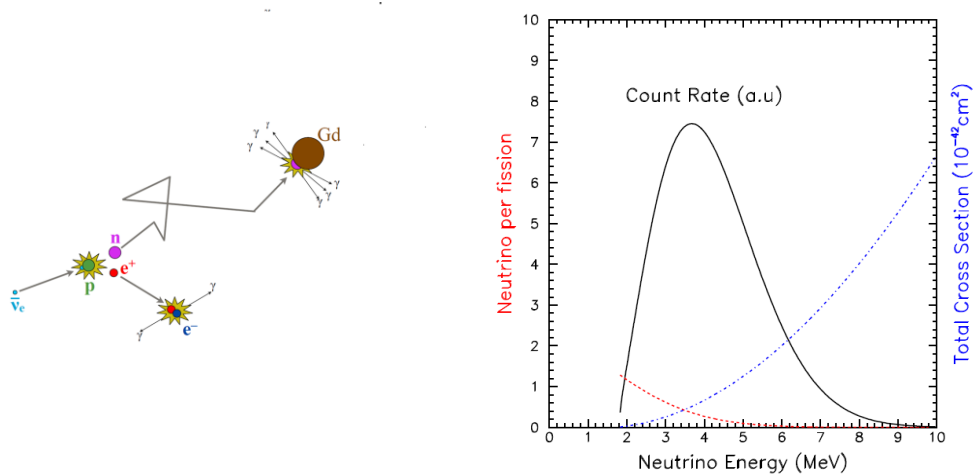

Figure 1: Left: Schematic depiction of an IBD event. Right: The shape of visible prompt energy spectrum as a convolution of the neutrino energy spectrum (using scale on left) and IBD cross section (using scale on right).

### 2.1.1. Only relative measurements are made

The neutrino flux from the reactors has an uncertainty of about 2-3%, so only a relative measurement, independent of the absolute neutrino flux, can reveal a small $\theta_{13}$. This requires at least one neutrino detector placed near the reactor to measure the flux of neutrinos, and a second detector placed at the far site to measure the oscillations. The disappearance of neutrinos can be deduced from the ratio between the far and near rates where most of the systematics cancel.

$$\frac{N_f}{N_n} = \left(\frac{N_{p,f}}{N_{p,n}}\right)\left(\frac{L_n}{L_f}\right)^2\left(\frac{\varepsilon_f}{\varepsilon_n}\right)\left[\frac{P_{\text{sur}}(E,L_f)}{P_{\text{sur}}(E,L_n)}\right] \quad (2)$$

In Equation 2, the subscripts f and n refer to quantities in the far and near sites, respectively. The survival probability ($P_{\text{sur}}$) is given by Equation 1.

### 2.1.2. Identical near and far detectors

In order to achieve a small systematic error in the ratio between the near and far detectors, the neutrino detectors are designed and built as identically as possible. The detectors are constructed as nested concentric cylindrical acrylic tanks, and filled with Gd-doped liquid scintillator, undoped liquid scintillator and mineral oil from inside out. The liquid scintillator is hydrocarbon solvent (Linear Alkyl Benzene (LAB) for Daya Bay and RENO and an o-PXE/n-dodecane mixture for DC) mixed with standard fluor (PPO) and wavelength shifter. The innermost tank is the target while the middle cylinder (the γ catcher) serves to detect gammas escaping the target volume. The outermost tank shields the target volumes from radioactivity of detector materials. For the Daya Bay experiment, the far and near detectors are composed of 4 and 2 identical modules, respectively, each with 20-ton target mass. The modules are constructed and filled as pairs to minimize systematic differences. The pairwise deployment allows side-by-side comparison. The modules in Daya Bay are shown to be functionally identical [11]. RENO has 1 near and 1 far detector of target mass listed in Table 1. The near detector for DC is still under construction at the time of this talk.

**2.1.3. Mitigation of backgrounds**

The backgrounds to $\theta_{13}$ measurement are well known from previous experiments. The gammas and neutrons produced by comic muons are potentially detrimental to a precision measurement. The far detectors are deployed underground with overburden of at least 300 mwe. To combat residual cosmogenic backgrounds, the detectors are enclosed in a steel tank, and submerged in water to attenuate ambient background from the surrounding rock and cosmogenic background by several orders of magnitude. The exact level of background reduction depends on the overburden and thickness of water shield, listed in Table 1. To eliminate fake IBD events generated by muons passing through the detector, the water pool is instrumented and covered by a muon detector. IBD events within some short interval of the muons are vetoed. Due to the low rate of residual muons in the far site, the deadtime due to the muon veto is small. The three experiments employed slightly different technologies for the muon veto. They are listed in Table 1.

Table 1. Key parameters of reactor experiments

| **Experiment** | **Daya Bay** | **Double Chooz (DC)** | **RENO** |
|---|---|---|---|
| Reactor configuration | 3×2 cores | 1×2 cores | 6×1 cores, inline |
| Detector configuration | 2 N + 1 F | 1 N + 1 F | 1 N +1F |
| Baseline (meter) | (364, 480, 1912) | (400, 1050) | (290,1380) |
| Overburden of far detector (mwe) | 860 | 300 | 450 |
| Detector geometry | Concentric cylinders of GdLS, γ-catcher and Oil buffer | | |
| Target mass (ton) | (40, 40, 80) | (10, 10) | (16.5, 16.5) |
| Outer shield | 2.5 m water | 0.50 m LS + 0.15 m steel | 1.5 of water |
| Muon veto | Water Cerenkov + RPC | LS + Scintillator Strips | Water Cerenkov |

## 2.2. Background

The analysis compares the IBD rates seen in the far and near detectors, and $\theta_{13}$ is extracted from Equation 2. The baselines are accurately measured by GPS and/or survey, typically to a few mm. The yields have to be corrected for background and relative efficiencies. While the ratio of near to far efficiencies is known to high precision, the backgrounds for the far and near sites have to be evaluated site by site. Below is a description of the main background contaminations, using Daya Bay numbers for illustration.

**2.2.1. Accidentals**

The random coincidence of prompt-like signals and neutron-like singles constitutes a major background to the IBD sample. The prompt-like and neutron-like signals are not caused by IBD events, but satisfy the event selection criteria by chance. This accidental background in Daya Bay is estimated from the neutron-like singles rate not correlated with IBD events from the data. The accidentals are the largest source of background in Day Bay, of order 2 and 5 % of the IBD rates in the near and far sites, respectively. The error of this background is only statistical, contributing less than 0.05% to the measurement error of Day Bay. Similar levels of accidental background are seen in DC and RENO.

**2.2.2. Fast neutrons**

Fast neutrons produced by cosmic muons can generate background by first interacting with the scintillator and being subsequently captured by Gd. The rate of these events in Daya Bay was determined from the data. The prompt events with energies not consistent with IBD events (10-100 MeV) are mostly due to fast neutrons which can be used to estimate the amount in the IBD signal region. This is done by fitting the energy spectrum of prompt events between 10 and 100 MeV, and extrapolating the distribution to the energy region of the IBD events. The estimated background is quite small for Daya Bay, order 0.2% of IBD events at all sites, but a conservative uncertainty, typically 30-50%, is assigned to this background.

**2.2.3. Li-9/He-8 background**

Long-lived isotopes produced by comic ray muons in the scintillator are not efficiently vetoed by the muon veto cut because of their long lifetimes. The most important contributors are Li-9 and He-8 with 178 and 119 ms half-lives, respectively. The beta decays of these isotopes mimic the prompt signal of an IBD event, whereas the excited parent sometimes emits a neutron in their de-excitation, faking the delayed neutron capture signal. The rate of these events can be estimated by studying the time distribution of signals after the passage of a muon. Again, the rate is small compared to that of the IBD in Daya Bay, but the uncertainty is large due to various unknowns in this process. This background is one of the main uncorrelated uncertainties in RENO and DC, order 1% at the far site, due to their shallower depth.

**2.2.4. Other backgrounds**

There are two other sources of background, one due to α-decays of residual radioactivity in the detector, common to all experiments, and another due to the Am-C sources, specific to Daya Bay. They are small, but assigned large uncertainty.

## 3. RESULTS

The analyses and results from each experiment are described below. Their unique features are pointed out.

### 3.1. Daya Bay

In order to perform a rate comparison, the yields at the three halls are background subtracted, and corrected for efficiency. The efficiencies of cuts applied are determined by data and Geant4 based simulation. Using the rates measured at the two near sites, the rate at the far site assuming no oscillation is predicted. Since there are 2 near sites in Daya Bay which has three pairs of reactors (a total of 6 cores), a weighted sum of the rates at the 2 near detectors is used for near site normalization. The observed rate at the far site was found to be about 5.4 % below non-oscillation expectation, indicative of oscillation effects. By fitting the observed rates at various sites to the oscillation hypothesis (see Figure 1, lower left) using uncorrelated systematic errors, Daya Bay arrived at a $\theta_{13}$ value of

$$\sin^2 2\theta_{13} = 0.089 \pm 0.010 \text{ (stat)} \pm 0.005 \text{ (syst)}.$$

The best fit and the prompt energy spectra at the near and far sites are shown in Figure 2 (see caption for details). The distortion of the far site prompt energy spectrum, while not used to determine $\theta_{13}$, supports the oscillation hypothesis.

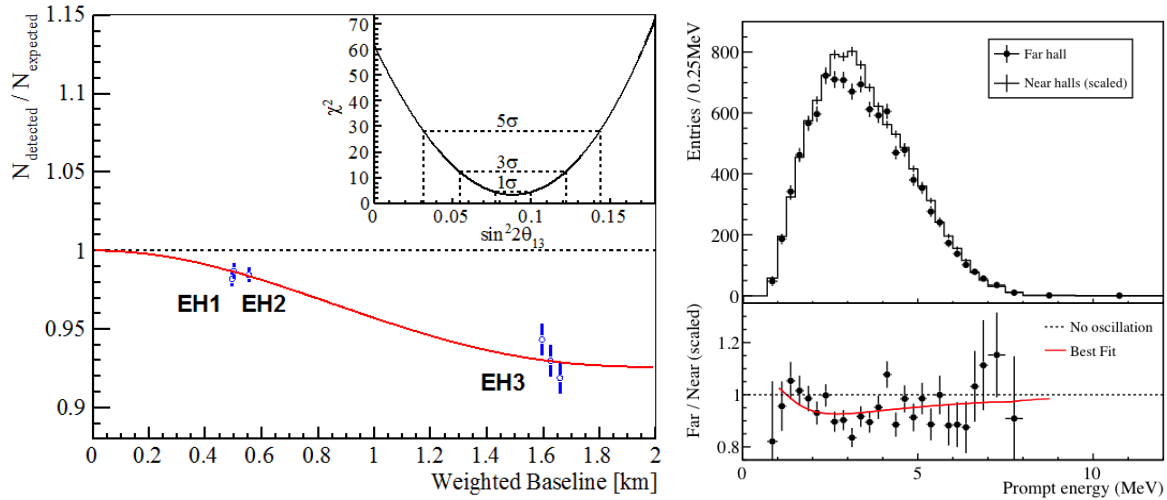

Figure 2: Daya Bay $\theta_{13}$ results. Left: Ratio of measured versus expected signals in each detector, assuming no oscillation. The error bar is the uncorrelated uncertainty of each detector, including statistical, detector-related, and background-related uncertainties. The expected signal has been corrected with the best-fit normalization parameter. Reactor and survey data were used to compute the flux-weighted average baselines. The oscillation survival probability at the best-fit value is given by the red (solid) curve. The dashed line is the no-oscillation prediction. The $\chi^2$ value versus $\sin^2 2\theta_{13}$ is shown in the inset. Right: The top panel is the measured prompt energy spectrum of the far hall compared with the no-oscillation prediction based on the measurements of the two near halls. Spectra were background subtracted. Uncertainties are statistical only. The bottom panel is the ratio of measured and predicted no-oscillation spectra. The red (solid) curve is the expected ratio with oscillations, calculated as a function of neutrino energy assuming $\sin^2 2\theta_{13} = 0.089$ obtained from the rate-based analysis. The dashed line is the no-oscillation prediction.

### 3.2. RENO

RENO employs an analysis procedure similar to that of Daya Bay. Using the rate measured in the near detector, the rate at the far site assuming no oscillation is predicted. A $\chi^2$ fit, using uncorrelated errors, was used to obtain the best value for $\theta_{13}$. The best-fit value for $\theta_{13}$ from RENO is

$$\sin^2 2\theta_{13} = 0.113 \pm 0.013 \,(\text{stat}) \pm 0.019 \,(\text{syst}).$$

The best fit and the prompt energy spectra at the near and far sites are shown in Figure 3 (see caption for details).

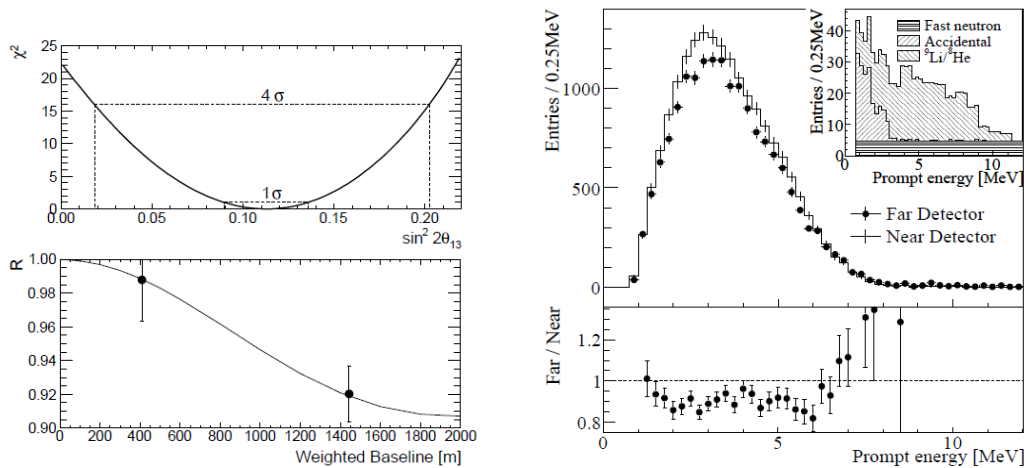

Insert PSN Here

Figure 3: RENO $\theta_{13}$ results. Left top: The χ distribution as a function of $\sin^2 2\theta_{13}$. Left bottom: Ratio of the measured reactor neutrino events relative to the expected with no oscillation. The curve represents theoretical oscillation survival probability at the best fit, as a function of the flux-weighted baselines. Right top: Observed spectrum of the prompt signals in the far detector compared with the non-oscillation predictions from the measurements in the near detector. The backgrounds shown in the inset are subtracted for the far spectrum. The background fraction is 5.5% (2.7%) for far (near) detector. Errors are statistical uncertainties only. Right bottom: The ratio of the measured spectrum of far detector to the non-oscillation prediction.

### 3.3. Double Chooz

Since the near detector at DC is still under construction at the time of this talk, the analysis for DC is based on calculated neutrino flux and absolute efficiencies. The systematic error in DC is dominated by the flux uncertainly. By normalizing the flux to measurement in Bugey-4 at 15 m [12], this uncertainty is reduced to 1.7%. Since publishing the first hint of non-zero value of $\theta_{13}$ based on rate and shape analyses of Gd-capture data [6], DC published an analysis of hydrogen capture data to obtain $\theta_{13}$ with comparable error [13]. The two DC $\theta_{13}$ results are:

$$\sin^2 2\theta_{13} = 0.109 \pm 0.030 \,(\text{stat}) \pm 0.025 \,(\text{syst}) \qquad (\text{Gd capture})$$
$$\sin^2 2\theta_{13} = 0.097 \pm 0.034 \,(\text{stat}) \pm 0.034 \,(\text{syst}) \qquad (\text{H capture})$$

The prompt energy spectra for Gd and H capture data, shown in Figure 4, are consistent with standard neutrino disappearance hypothesis (see caption for details).

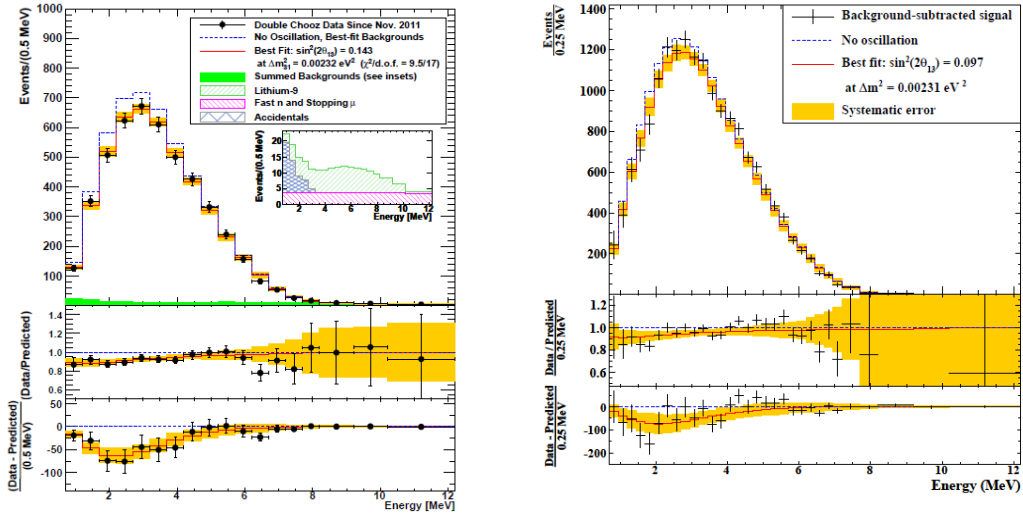

Figure 4: DC $\theta_{13}$ results. Left top: Measured prompt energy spectrum superimposed on the expected prompt energy spectrum for Gd capture data, including backgrounds (green region), for the no-oscillation (blue dotted curve) and best-fit (red solid curve) at $\sin^2 2\theta_{13} = 0.109$ and $\Delta m^2_{31} = 2.32 \times 10^{-3}$ eV$^2$. Inset: stacked spectra of backgrounds. Left bottom: differences between data and no-oscillation prediction (data points), and differences between best fit prediction and no-oscillation prediction (red curve). The gold band represents the systematic uncertainties on the best-fit prediction. Right: Same as those of left for hydrogen capture data.

### 4. SUMMARY

The recently published $\theta_{13}$ results from Daya Bay, RENO and DC are summarized. A nonzero value of $\theta_{13} \sim 9°$ is firmly established. This removes one of the unknowns in the neutrino sector. $\theta_{13}$ is now measured with a fractional error of about 10%, and significant improvement in precision is expected in the next few years from reactor experiments. The relatively large and precise value of $\theta_{13}$ has already begun to provide non-trivial implications on the parameters of the 3-flavor paradigm [14]. More ramifications on future neutrino and related programs are expected to come.

## 5. ACKNOWLEDGMENTS

I thank my collaborators at Daya Bay for helpful discussions and for providing useful information about the Day Bay experiment. This work is supported by US Department of Energy contract DE-FG02-07ER41518.